\documentstyle[PASJadd]{PASJ95}
\draft
\newcommand{\vol}{}
\newcommand{\no}{}
\newcommand{\lett}{}
\newcommand{\spage}{}
\newcommand{\rdate}{}
\newcommand{\adate}{}
\begin{document}

\title{Elliptical Lens Model in PG 1115+080}
\author{Katsuaki {\sc Asano} and Takeshi {\sc Fukuyama} \\
{\it Department of Physics, Ritsumeikan University,
	Kusatsu, Shiga 525-8577} \\
{\it E-mail(KA): sph10001@bkc.ritsumei.ac.jp}}

\abst{ 
The mass distribution of the lensing galaxy of the
multiple quasar PG1115+080 has been studied.
Using the observational data of Christian et al.
(1987, ApJ 464, 92),
we applied a single elliptical lens model
with a softened power-law behavior.
It has been revealed that the image positions, amplifications and
time delays of this lensing system cannot be explained
by the single transparent lens model.
We compare this elliptical lens model with a multipole
expansion model.
}
\kword{Gravitational lenses---Galaxies: structure---Quasars:
individual(PG 1115+080)}
\maketitle
\thispagestyle{headings}


\newpage
\section{Introduction}

We discuss the multiple quasar PG 1115+080 using an elliptical lens model
with a softend power-law behavior.

The same theme was discussed by Narasimha et al. (1982).
The motivation to repeat this research was as follows.
Firstly, they used the observational data by Young et al. (1981).
However, the more recent data by Christian et al. (1987)
have been adopted by the references Yoshida and Omote (1988),
Kakigi et al. (1995).
Our purpose, however, was not simply to reanalyze the same object using the new
observational data, but to discuss gravitational lensing from many-sided view
points.
Let us explain its physical meaning more concretely.
In gravitational lensing,
``the number of degrees of freedom'' ($\equiv$ number
of fitted observed data$-$
number of fitting parameters,
which we call DOF hereafter)
is small and it is dangerous to deduce any conclusion
by simply having reproduced, for instance,
image positions.
Indeed, many qualitatively different models can reproduce
the restricted observed data.
However, this view point has so far gone unnoticed, and usually
a numerically tractable simple model has been applied to the respective
lensing phenomenon.
It is necessary for us to apply various lensing models
systematically to the same lensing phenomenon and to compare
their results with each other.
Together with these numerical approaches,
analytical surveys of lensing models are also needed
for obtaining a systematic understanding of lensing phenomena.

In this paper, along this line of thought, we systematically
formulate an elliptical
mass distribution model with various softened power-law behavior.
We then apply them to the multiple quasar PG 1115+080.
The results are compared with those obtained by a qualitatively
different model, a multipole expansion model.

This lensing galaxy is relatively isolated in comparison with
other lensing events, and has quadruple images of the background
quasars (Young et al. 1981; Christian et al. 1987; Kristian et al. 1993).
Therefore, this system may be suitable for applying an
analytical treatment of the lens models developed in this paper.
However, the physical parameters of the lensing object
are found to be sensitive
to the observed image positions, 
and especially to the lens position
which have been measured with some statistical
treatments.
In order to obtain reliable pictures of the lensing object we need 
a variety of more
precise data of images.
They are the image shape and the precise spectroscopies of
every image component and their time variations etc.
Together with these improvements of fitted data,
it is also necessary to apply qualitatively
different models to the same lensing object.
For instance, although the elliptical lens model
is transparent, the multipole expansion
model (Kakigi et al. 1995; Fukuyama et al. 1997)
is non-transparent.
They give various and even inconsistent aspects of the lensing object.
However, from such an variety of aspects, we expect that the true features
of the lensing object will be uncovered.
This article is a step towards this goal.

The present paper is organized as follows.
We give a general formulation of the elliptical lens model with
a softened power-law behavior in section 2.
The case for $k \leq 3$ (power of distance dependence on the lensing mass
density) is discussed in detail.
In section 3, the given formulation is applied to the
PG 1115+080 lensing system.
The best-fit parameters for $k=2$ and $k=3$ are given.
These parameters give the aspect of a lensing object
that is qualitatively
different from that of Narasimha et al.
This mainly comes from the fact that the lens position
assumed by Narashimha et al. is different from ours, which is
fixed with the observed position by Christian et al.
In section 4, we argue about the discrepancies of our results
with the observations
and compare the elliptical lens model with the multipole expansion
model.
Very recently, lens models of PG 1115+080 accompanying
the group galaxies have been applied
by Keeton and Kochanek (1997) and Schechter et al. (1997).
We make some comments concerning them in relation to our article.

\section{Formulation}
Our lens model is described based on the mass density of Bourassa et al.
(1973)
(see also
Bourassa, Kantowski 1975; Schramm 1990; Kormann et al. 1994;
Kassiola, Kovner 1993),
\begin{equation}
\rho (a)=
\left\{ \begin{array}{ll}
\rho_0 \left\{1+{\displaystyle \left(\frac{a}{r_{\rm c}}\right)^2}\right\}
^{-\frac{k}{2}},&
\mbox{for $a \leq n r_{\rm c}$}\\
0, & \mbox{for $a > n r_{\rm c}$} 
\end{array} \right. 
\end{equation}
with
\begin{equation}
x'^2+y'^2+{z'^2 \over 1-e^2} \equiv a^2.
\label{a2}
\end{equation}
Here, $r_{\rm c}$
is the core radius and $\rho_0$ is the constant central density;
$e$ is the eccentricity.

We choose the direction of the line-of-sight on the $z$-axis,
which is tilted relative
to the symmetry axis $z'$ by angle $\gamma$.
The situation is depicted by figure 1.

\begin{center}
  ----------  \\
    Figure1     \\
  ----------
\end{center}
Although $\phi$ is a rotation angle between the x-axis and the major axis
of the lens object,
in this section we assume $\phi=0$.
The principal coordinates of a spheroid $(x',y',z')$ are related with $(x,y,z)$
by 

\begin{equation}
\pmatrix{x'\cr y'\cr z'\cr}=\pmatrix{1&0&0\cr 0&\cos{\gamma}&{-\sin{\gamma}}\cr
0&\sin{\gamma}&\cos{\gamma}\cr} \pmatrix{x\cr y\cr z\cr}.
\label{gamma}
\end{equation}
Substituting equation (\ref{gamma}) into equation (\ref{a2}), we obtain
\begin{equation}
a^2=b^2+{1-e^2 \sin^2{\gamma} \over 1-e^2} \left(z+{e^2 y \cos{\gamma}
\sin{\gamma}
\over 1-e^2 \sin^2{\gamma}}\right)^2,
\end{equation}
where
\begin{equation}
b^2 \equiv x^2+{y^2 \over 1-e^2 \sin^2{\gamma}}.
\label{b2}
\end{equation}
The surface mass density, $\Sigma (x,y) \equiv \smallint dz \rho (a)$,
is given by
\begin{equation}
\Sigma (x,y)=\sqrt{1-e^2 \over 1-e^2 \sin^2 {\gamma}} \int_{b^2}^{(n
r_{\rm c})^2} da^2
{\rho(a) \over \sqrt{a^2-b^2}} \equiv
\sqrt{1-e^2 \over 1-e^2 \sin^2{\gamma}} K_k.
\label{sigma}
\end{equation}
Here, $K_k$ have the following forms, depending on the parameter $k$:
\begin{eqnarray}
K_1&=&2 \rho_0 r_{\rm c} \ln{\sqrt{n^2+1}+\sqrt{n^2-
(\frac{b}{r_{\rm c}})^2} \over 
\sqrt{1+\left(\frac{b}{r_{\rm c}}\right)^2}}, \\
K_2&=&\frac{2 \rho_0 r_{\rm c}}{\sqrt{1+(\frac{b}{r_{\rm c}})^2}}
\tan^{-1}{\sqrt{n^2-(\frac{b}{r_{\rm c}})^2} \over \sqrt{1+(\frac{b}
{r_{\rm c}})^2}}, 
\label{k=2} \\
K_3&=&\frac{2 \rho_0 r_{\rm c}}{1+(\frac{b}{r_{\rm c}})^2}
\frac{\sqrt{n^2-(\frac{b}{r_{\rm c}})^2}}{\sqrt{n^2+1}}, \\
K_5&=&\frac{2 \rho_0 r_{\rm c}}{(1+(\frac{b}{r_{\rm c}})^2)^2}
\frac{\sqrt{n^2-(\frac{b}{r_{\rm c}})^2}}{3 (n^2+1)^{3/2}}
(3+2 n^2+(\frac{b}{r_{\rm c}})^2).
\end{eqnarray}
Fermat's potential is given by
\begin{equation}
\phi (\mbox{\boldmath$x$},\mbox{\boldmath$x_{\rm s}$})
=\frac{1}{2} (\mbox{\boldmath$x$}-\mbox{\boldmath$x_{\rm s}$})^2
-\frac{1}{\pi} \int dx' dy' \kappa (\mbox{\boldmath$x'$})
\ln{\vert \mbox{\boldmath$x$}-\mbox{\boldmath$x_{\rm s}$} \vert}.
\label{Fermat}
\end{equation}
Here, $\mbox{\boldmath$x$}$ and $\mbox{\boldmath$x_{\rm s}$}$
are the two-dimensional image and
source position, respectively.
Also, $\kappa$ is the surface mass density of the lens object,
\begin{equation}
\kappa (\mbox{\boldmath$x$})
=\frac{\Sigma (\mbox{\boldmath$x$})}{\Sigma_{\rm crit}},
\end{equation}
normalized by 
\begin{equation}
\Sigma_{\rm crit} \equiv \frac{c^2 D_{\rm s}}{4 \pi G D_{\rm l} D_{\rm ls}}
(\equiv \frac{c^2}{4 \pi G D}).
\label{D}
\end{equation}
Notations are depicted in figure 2.

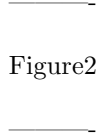
\begin{center}
  ----------  \\
    Figure2     \\
  ----------
\end{center}
The lens equation is
\begin{equation}
\mbox{\boldmath$x_{\rm s}$}=\mbox{\boldmath$x$}
-\frac{1}{\pi} \int dx' dy' \kappa (\mbox{\boldmath$x'$})
\frac{\mbox{\boldmath$x$}-\mbox{\boldmath$x'$}}{(\mbox{\boldmath$x$}-\mbox{\boldmath$x'$})^2}.
\end{equation}
In complex notation it becomes
\begin{equation}
z_{\rm s}=z-\frac{4 G D}{c^2} I^* (z),
\end{equation}
where $z_{\rm s} \equiv x_{\rm s}+i y_{\rm s}$ and $z \equiv x+iy$
(here and hereafter we use the letter $z$ in this sense).
$I^*(z)$ is the complex conjugate of $I(z)$,
which is defined by
\begin{equation}
I \equiv \int dx' dy' \Sigma (b') \frac{1}{z-z'}.
\end{equation}
Taking
\begin{equation}
x' =b' \cos{\varphi'}, \qquad y'=b'\sqrt{1-e^2 \sin^2{\gamma}} \sin{\varphi'}
\end{equation}
into consideration, $I$ is transformed to
\begin{equation}
I=\sqrt{1-e^2  \sin^2{\gamma}} \int_{0}^{n r_{\rm c}}
db' b' \Sigma (b') \oint d\varphi'
\frac{1}{z-b' (\cos{\varphi'} +i \sqrt{1-e^2 \sin^2{\gamma}} \sin{\varphi'})}.
\end{equation}
Here,
\begin{equation}
\oint d\varphi' \frac{1}{z-b' (\cos{\varphi'} +i \sqrt{1-e^2 \sin^2{\gamma}}
\sin{\varphi'})}=
\left\{ \begin{array}{ll}
\frac{\sqrt{z^2}}{z}\frac{2 \pi}{\sqrt{z^2-b'^2 e^2 \sin^2{\gamma}}},&
\mbox{for $b'^2 \leq b^2=x^2+\frac{y^2}{1-e^2 \sin^2{\gamma}}$}\\
0, & \mbox{for $b'^2 > b^2$}.
\end{array} \right.
\label{Newton}
\end{equation}
We then obtain
\begin{equation}
I_k=2 \pi \frac{\sqrt{z^2}}{z} \sqrt{1-e^2 \sin^2{\gamma}}
\int_{0}^{{\rm min}(n r_{\rm c},b)} db' \frac{b' \Sigma_k (b')}{\sqrt{z^2-b'^2 e^2 \sin^2{\gamma}}}.
\label{Ik}
\end{equation}
We consider the $k=2,3$ cases of mass density,
the motivation of which
is as follows.
The observed brightness distribution of galaxies decays
as $r^{-3}$ (Binney, Tremaine 1987).
The mass-to-light ratio suggests that the mass distribution
decays more mildly than that.

For $k=3$, $I_3$ takes the following form:
\begin{equation}
I_3=\frac{2 \pi \rho_0 r_{\rm c}^2}{\sqrt{n^2+1}}
\frac{\sqrt{z^2}}{z} \frac{\sqrt{1-e^2}}{e \sin{\gamma}}
\int_{0}^{{\rm min}(n^2,(\frac{b}{r_{\rm c}})^2)}
\frac{d\omega}{1+\omega} \sqrt{\frac{\omega-n^2}
{w-(\frac{z}{r_{\rm c} e \sin{\gamma}})^2}}.
\end{equation}
Here, we consider the following indefinite integral:
\begin{eqnarray}
F &\equiv& \int \frac{d\omega}{1+\omega} \sqrt{\frac{\omega-\beta}{\omega-\alpha}},
\qquad \alpha \in {\bf C}, \beta \in {\bf R} \nonumber \\
&=&\sqrt{\frac{1+\beta}{1+\alpha}}
\ln{\frac{\sqrt{(1+\alpha)(\omega-\beta)}
-\sqrt{(1+\beta)(\omega-\alpha)}}
{\sqrt{(1+\alpha)(\omega-\beta)}+\sqrt{(1+\beta)(\omega-\alpha)}}}
\nonumber \\
& &-\ln{\frac{\sqrt{\omega-\beta}-\sqrt{\omega-\alpha}}
{\sqrt{\omega-\beta}+\sqrt{\omega-\alpha}}}.
\end{eqnarray}
Thus, if the images are positioned inside of the cut-off radius $n r_{\rm c}$,
\begin{eqnarray}
I_3=2 \pi \rho_0 r_{\rm c} \frac{\sqrt{z^2}}{z} \frac{\sqrt{1-e^2}}{e \sin{\gamma}}
\left[ \frac{1}{\sqrt{n^2+1}}\ln{\frac{(z_0-n e \sin{\gamma})
(A_0+\sqrt{n^2-B_0^2})}{(z_0+n e \sin{\gamma})(A_0-\sqrt{n^2-B_0^2})}}
\right. \nonumber \\
+\frac{e \sin{\gamma}}{\sqrt{z_0^2+e^2 \sin^2{\gamma}}}
\left\{
\ln{\frac{z_0 \sqrt{n^2+1}+n \sqrt{z_0^2+e^2 \sin^2{\gamma}}}
{z_0 \sqrt{n^2+1}-n \sqrt{z_0^2+e^2 \sin^2{\gamma}}}}
\right. \nonumber \\
\left. \left.
-\ln{\frac{\sqrt{n^2+1} A_0 e \sin{\gamma}
+\sqrt{(z_0^2+e^2 \sin^2{\gamma})
(n^2-B_0^2)}}
{\sqrt{n^2+1} A_0 e \sin{\gamma}-\sqrt{(z_0^2+e^2 \sin^2{\gamma})(n^2-B_0^2)}}
}
\right\} \right],
\end{eqnarray}
where
\begin{eqnarray}
A_0 \equiv \frac{\sqrt{1-e^2 \sin^2{\gamma}}}{e r_{\rm c} \sin{\gamma}}
(x+\frac{i y}{1-e^2 \sin^2{\gamma}}),
\qquad
B_0 \equiv \frac{b}{r_{\rm c}}
\end{eqnarray}
and
\begin{eqnarray}
z_0 \equiv \frac{z}{r_{\rm c}}.
\end{eqnarray}
However, if the images are located outside of the cut-off radius, $I_3$ takes
the following form:
\begin{eqnarray}
I_3=2 \pi \rho_0 r_{\rm c} \frac{\sqrt{z^2}}{z} \frac{\sqrt{1-e^2}}{e \sin{\gamma}}
\left[ \frac{1}{\sqrt{n^2+1}}\ln{\frac{(z_0-n e \sin{\gamma})}
{(z_0+n e \sin{\gamma})}}
\right. \nonumber \\
\left. +\frac{e \sin{\gamma}}{\sqrt{z_0^2+e^2 \sin^2{\gamma}}}
\ln{\frac{z_0 \sqrt{n^2+1}+n \sqrt{z_0^2+e^2 \sin^2{\gamma}}}
{z_0 \sqrt{n^2+1}-n \sqrt{z_0^2+e^2 \sin^2{\gamma}}}} \right].
\end{eqnarray}
In the case of $k=2$, from equations (\ref{k=2}) and (\ref{Ik}) it follows
that
\begin{equation}
I_2=2 \pi \rho_0 r_{\rm c}^2 \sqrt{1-e^2} \frac{\sqrt{z^2}}{z}
\int_{0}^{{\rm min}(n^2,B_0^2)} d\omega \frac{1}{\sqrt{1+\omega}}
\tan^{-1}{\sqrt{\frac{n^2-\omega}{1+\omega}}} \cdot
\frac{1}{\sqrt{z_0^2-\omega^2 e^2 \sin^2{\gamma}}}.
\label{I2}
\end{equation}
Equation (\ref{I2}) can not be solved analytically, except for $n=\infty$.
As for $k=2$, there is a more conventional model which is used to fit the
X-ray surface brightness profile (Miralda-Escud\'e 1993):
\begin{equation}
\kappa(\theta,\phi)=\left\{ \begin{array}{ll}
\frac{b_0}{2 \sqrt{\theta^2 (1-\epsilon \cos{2 \phi})+r_{\rm c}^2}},
&  \mbox{for $\theta \leq \theta_{\rm cut}$}\\
0, & \mbox{for $\theta > \theta_{\rm cut}$.} 
\end{array} \right. 
\label{x2}
\end{equation}
Here, $b_0=2 \pi \rho_0 r_{\rm c}^2$, and $\epsilon$ is the ellipticity
related to the eccentricity by
\begin{equation}
\epsilon = \frac{e^2}{2-e^2}.
\end{equation}
As can be easily understood from equations (\ref{sigma}) and (\ref{k=2}),
this corresponds to the $n \to \infty$ limit in equations
(\ref{sigma}),(\ref{k=2})
with $\gamma=\pi/2$.
The length scale in equation (\ref{x2}) is $1/\sqrt{1-\epsilon}$ times longer
than that in equation (\ref{I2}).
A cut-off of mass distribution is introduced at the level of
the surface mass density.
Equation (\ref{I2}) seems to be physically more general
than equation (\ref{x2}), because
equation (\ref{x2}) has no cut-off radius and we can use equation (\ref{I2})
with $\gamma \neq \pi/2$.
We thus consider equation (\ref{I2}).
In order to calculate equation (\ref{I2}), we expand it with respect to
$e^2 \sin^2{\gamma}$.
Using
\begin{equation}
\frac{1}{\sqrt{z_0^2-\omega e^2 \sin^2{\gamma}}}
\simeq \frac{1}{z_0} + \frac{\omega}{2 z_0^3} e^2 \sin^2{\gamma}
+\frac{3 \omega^2}{8 z_0^5} e^4 \sin^4{\gamma},
\label{app}
\end{equation}
$I_2$ becomes
\begin{equation}
I_2=2 \pi \rho_0 r_{\rm c}^2 \sqrt{1-e^2} \frac{\sqrt{z^2}}{z}
\left( \frac{1}{z_0} Q_0+\frac{1}{2 z_0^3} e^2 \sin^2{\gamma} Q_1
+\frac{3}{8 z_0^5} e^4 \sin^4{\gamma} Q_2 \right).
\end{equation}
Here, $Q_n$ is defined by
\begin{equation}
Q_n \equiv \int_{0}^{{\rm min}(n^2,B_0^2)} \frac{d\omega}{\sqrt{1+\omega}}
\tan^{-1}{\sqrt{\frac{n^2-\omega}{1+\omega}}} \omega^n .
\end{equation}
Their explicit forms are
\begin{eqnarray}
Q_0 = \left[ 2 \sqrt{1+\omega} \tan^{-1}{\sqrt{\frac{n^2-\omega}{1+\omega}}}
-2 \sqrt{n^2-\omega} \right]_{0}^{{\rm min}(n^2,B_0^2)}, ~\qquad \qquad
\qquad \qquad \qquad \qquad \; \, \\
Q_1 = \left[ \frac{1}{9} (12-4 n^2-2 \omega) \sqrt{n^2-\omega}
+\frac{2}{3} (\omega-2) \sqrt{1+\omega}
\tan^{-1}{\sqrt{\frac{n^2-\omega}{1+\omega}}}
\right]_{0}^{{\rm min}(n^2,B_0^2)}, ~~~~~~\,\\
Q_2 = \left[ \frac{\sqrt{n^2-\omega}}{225} \left(
-16 (3 n^4-5 n^2+15)+8 (5-3 n^2) \omega-18 \omega^2 \right)~~~~~
\qquad \qquad \; \; \right. \qquad \qquad \nonumber \\
\left.  -\frac{2}{15} (3 \omega^2-4 \omega+8) \sqrt{1+\omega}
\tan^{-1}{\sqrt{\frac{n^2-\omega}{1+\omega}}}
\right]_{0}^{{\rm min}(n^2,B_0^2)}.
\end{eqnarray}

\section{Numerical Calculation of Multiple Images of PG 1115+080}

In this section we apply the elliptical lens model with $k=2$ and $k=3$
to the multiple quasar of PG 1115+080.
The number of model parameters is 7, namely $n$, $e \sin{\gamma}$,
$r_{\rm c}$, $\rho_0 \sqrt{1-e^2}$, $\phi$, and source positions.

Here, we have introduced the dimensionless parameter $\kappa_0$ in place
of $\rho_0$, defined by
\begin{equation}
\kappa_0 \equiv
\frac{8 \pi G D_{\rm l} D_{\rm ls}}{D_{\rm s}} \rho_0 r_{\rm c},
\end{equation}
which appears as a common factor of the normalized surface mass 
density.
Cosmological parameters, such as $z_{\rm l}$, $z_{\rm s}$, the
Hubble constant and
$\Omega_0$, are assumed to be given.
The mass of the lensing galaxy is expressed from equation (\ref{sigma}) as
\begin{eqnarray}
M_k=\int \Sigma_k dx dy=\sqrt{\frac{1-e^2}{1-e^2 \sin^2{\gamma}}}
\int K_k dx dy,
\end{eqnarray}
and calculated to be
\begin{equation}
M_k=4 \pi r_{\rm c}^3 \rho_0 \sqrt{1-e^2} \delta_k.
\end{equation}
Here, $\delta_k$ are given by
\begin{equation}
\delta_3 \equiv \frac{1}{2 \sqrt{1+n^2}} \int_{0}^{n^2}
\frac{\sqrt{n^2-\omega}}{\sqrt{1+\omega}}
=\ln{(\sqrt{1+n^2}+n)}-\frac{n}{\sqrt{1+n^2}}
\end{equation}
and
\begin{equation}
\delta_2 \equiv \frac{1}{2} \int_{0}^{n^2} \frac{1}{\sqrt{1+\omega}}
\tan^{-1}{\sqrt{\frac{n^2-\omega}{1+\omega}}} d\omega
=n-\tan^{-1}{n}
\end{equation}
The total number of observed data of Christian et al.
is 11, i.e. $4 \times 2$
for image positions and 3 for the flux ratios of images (table 2).

We have adopted a new datum for the redshift of the lensing
object, $z_{\rm l}=0.294$, from Angonin-Willaime et al. (1993).
Our fitting procedures were as follows.
We obtained parameter values by minimizing
the residuals in the source plane, which is defined by
\begin{equation}
f^2 = \sum_{i=1}^4 \left| z_{\rm s}-z_i+\frac{4 G D}{c^2} I^*(z_i)
\right|^2.
\label{chi}
\end{equation}
Here $z_i$ is the $i$-th image position in a complex representation,
and we substituted the observed value into it.
The image positions are given by the following radial coordinates:
\begin{eqnarray}
l_i=\sqrt{x_i^2+y_i^2}, \qquad \theta_i=\tan^{-1}{\frac{y_i}{x_i}}.
\end{eqnarray}
In section 2, we supposed $\phi=0$.
In general, however, the major axis does not coincide
with the $x$-axis (we adopted it parallel to right ascension).
Taking the rotation $\phi$ in the lens surface into consideration,
$z_i$ in equation (\ref{chi}) is given by
\begin{equation}
z_i=l_i \cos{(\theta_i+\phi)}+i l_i \sin{(\theta_i+\phi)}.
\end{equation}
In order to find the minimum for the $f^2$, we have used
the downhill-simplex method.
We have not incorporated any amplification data in equation (\ref{chi})
because of their unreliability compared with the image positions.
We also have not considered the weight in each image,
since the error estimate of each image was not given.
We calculated the least square function for the $k=2$
and $k=3$ models.
In order to confirm our calculation,
it was checked whether our numerical calculations could reproduce
the results of Narasimha et al.
if we used the same data as theirs.

We had many parameters, and the fitting parameter region was not
necessarily so restricted.
Thus, we first fix two parameters, $n$ and $e\sin{\gamma}$, and then 5
fitting parameters remained to be adjusted.
We then changed $e\sin{\gamma}$  from 0 to 1 by a 0.1 interval.
We also chose $n=5, 10, 20, 30$.
For $k$=3 we could fit the parameters in all ranges of $n$;
the other parameters were insensitive to the $n$ value.
On the other hand, for $k$=2 we could fit only for $n=5$ and 10.
For $k$=3, the surface mass density is almost constant
for $n \geq 5$,
but for $k$=2 it continues to increase for $n \geq 5$.
In these procedures, we did not intend to uniquely fit all of
the parameters,
but to consider the best fitting under the given physically probable
parameters, $e \sin{\gamma}$, $n$.
Using these parameters we also calculated the image positions,
amplifications and time delays etc.
The results of the best fitting are listed in tables 3 and 4.
Table 3 (table 4) is the model for $k=2$ ($k=3$).
The fitted $\phi$'s are very close to each other
in spite of the difference in the other parameters.
In these tables we have listed the observed image positions in
the coordinates, where the $x$-axis
coincides with the major axis of the lenses.
The observed image positions and their numerical calculations are depicted
in figure 3.

\begin{center}
  ----------  \\
    Figure3     \\
  ----------
\end{center}

From these tables there appears the following image of the lensing object.
The core radius ranges over $3$-$5 h^{-1}$ kpc.
These values are very large compared with the generally accepted core size.
Most of the observations and analyses suggest that the
core sizes of elliptical galaxies are few hundred pc
(Lauer 1985; Lauer et al. 1991; Bertin et al. 1993).
Thus, this large core radius places
the single elliptical lens model at a disadvantage.
The total mass is roughly
$10^{12} \MO$ under the assumption $n=5$ or $10$.
The central density $\rho_0$ roughly equals $10^{-23} h^2$ g ${\rm cm^{-3}}$.
Therefore, our models exhibit a very weak central concentration compared
with the general results of analyses applied to the other
lensing events (Kent, Falco 1988; Grogin, Narayan 1996).
We have obtained a good fitting for $e \sin{\gamma}=0.3$-$0.6$.
$e \sin{\gamma}$ (apparent eccentricity) $=0.3$
($e \sin{\gamma}=0.6$) is regarded as an $E_0$ ($E_2$) galaxy.
Thus, our results are compatible with Schechter's observation
that the lensing galaxy is round to within $10\%$ (Schechter 1996).
The cut-off radius corresponds to $30$-$50$ Kpc for $n=10$.
These estimates depend on the cosmological
parameter $\Omega_0$.
It changes the scale of the parameters through $D$ of equation(\ref{D}).
If we assume $\Omega_0=0$ in place of $\Omega_0=1$,
the core radius,
total mass and $\rho_0$ are estimated to 1.08-, 1.08- and 0.86-times
larger than those for $\Omega_0=1$, respectively.

As for the image positions,
these parameters do not well reproduce the observed values correctly.
Especially, as $e\sin{\gamma}$ becomes larger,
it is difficult to fit both ${\rm A}_1$ and ${\rm A}_2$ with the observations.
In their observational data, Christian et al.
did not give an error estimate of each image.
However, Kristian et al. (1993) observed PG 1115 by HST
and determined the image positions at a 5 mas error.
These data, however, are not essentially different
from the data of Christian et al.
By using the data by Kristian et al.,
we obtained the same parameters as those given in tables 3 and 4 within
a 10\% difference.
We calculated the $\chi^2$ of the image positions from these
parameters and data.
For example, we obtained the values $\chi^2/{\rm DOF}=1383.48/3$
for $k=2$, $n=10$, and $e \sin{\gamma}=0.3$,
and $\chi^2/{\rm DOF}=572.64/3$
for $k=3$, $n=10$, and $e \sin{\gamma}=0.5$.
These values are extremely large ($\chi^2/{\rm DOF} \gg 1$).

As for the image amplifications, the coincidence of the calculated values with
the observed ones is not very good.
Especially, the calculated magnifications of the B and C components are in
reverse order with the observation.
Moreover, the predicted value of the undetected image D is too bright.

The predicted time delays in tables 3 and 4
are too short compared with the observation
$\Delta t_{\rm BC}=23$-$28$ days (Schechter et al.1997;
Bar-Kana 1997).
These results insist that a single transparent lens model
can not be applied to this lensing system without any perturbation.

However, we can see the tendency of the parameters
according to the change of other parameters for given observational
data.
From tables 3 and 4, the larger $e \sin{\gamma}$ becomes,
the smaller does the core radius appear.
The core radius for $k=2$ is about 30\% smaller than that for $k=3$.
On the contrary,
the central density becomes slightly larger,
but remains almost unchanged,
according to the decreasing core size.
The conclusion that the core size becomes smaller
with the power law $k$ was also reported
by Grogin and Narayan (1996) and Keeton and
Kochanek (1997).
The reason is as follows.
Any lens model predicts almost the same mass
within the ring of images.
If $k$ becomes smaller,
the mass becomes larger.
Therefore, in order to decrease the mass,
the core radius must be small with a constant central density.
We also remark that the core radius for $k=2$ decreases by about 10\%
when $n$ changes from 5 to 10,
while that for $k=3$ almost does not change.
Most all researchers have used the expressions
for $n=\infty$,
because this simplification makes the calculation tractable.
However, we left $n$ finite and used approximation (\ref{app}).
The discrepancy between the behaviors of $k=2$
and $3$ may partly come from this difference.
In order to check this possibility,
we studied the $n$ dependence of the core radius
by a spherical symmetric model which is tractable
without any approximation.
The result is as follows.
For $n<10$, the core radius grows by the order of
1/100 sec totally as $n$ increases,
which is negligible in comparison
with the present precision of observation.
The core radius for $n>10$ is almost the same as for $n=\infty$.
However, we were forced to use $n<10$, since the core radius
is relatively large in our case.
The reason why the dependence of core radius on $n$
in our model is larger than the test calculation
in the spherical symmetric model may be due to
approximation (\ref{app}).

If the core radius is about several hundred pc,
as conventionally adopted,
we can take the $n$ order to be several hundred.
Thus, $n$ can not play any important role,
though $\rho_0$ decreases as $n$ increases.

Let us compare our result with that by Narasimha et al. for the same
$k=3$ model.
For example, some of their best-fit parameters are $n=20, e\sin{\gamma}=0.6,
r_{\rm c}=0^{\prime \prime}.27$, and $\kappa_0 \sqrt{1-e^2}=5.946$.
Therefore, their model exhibits a much more compact lens model than ours.
This discrepancy comes from the difference in the accepted observed
data.
In their article the lens position was left as a free parameter,
since the lensing galaxy had not yet been identified at that time.
Therefore, their method of parameter fitting is essentially different
from ours.
Also, they fit the separations between the images only with observation;
therefore, they neglected the relative position of the images and the lens.
The data which Narasimha et al. used are ${\rm AB}=1.9$,
${\rm BC}=2.4$, ${\rm CA}=2.0$ and ${\rm A}_1{\rm A}_2=0.54$ in arcsec.
On the other hand, the data of Christian et al. are
${\rm AB}=1.8$, ${\rm BC}=2.0$, ${\rm CA}=2.3$ and ${\rm A}_1{\rm A}_2=0.49$.
${\rm BC}$ and ${\rm CA}$ are reversed in two data.
Moreover, their results can not correctly reproduce the observed data.
They recorded the separations of images in the unit of the core radius.
In matching to the observed image separations, however,
their core radius does not converge on one value,
and disperses from $0^{\prime \prime}.26$
to $0^{\prime \prime}.37$ for each image separation.

Thus, the model parameters are sensitive to the image positions,
and especially to the lens position..
To be worse, the error bar of the lens position is still very large,
50 mas (Kristian et al. 1993).

\section{Discussion}

As we have discussed, the elliptical lens models commonly
used give no good fitting,
especially concerning the observed image amplifications.
Let us consider the possibilities that the observations are biased by
something not under consideration so far.
Image B is positioned nearer to the lensing galaxy than C
(and ${\rm A}_1$, ${\rm A}_2$).
Therefore,
B might be brighter than C, but would suffer absorption due to the atmosphere
of the lensing galaxy.
The old data (Young et al. 1981) suggested this possibility.
However, photometry of the QSO components in the new CCD frames (Christian
et al. 1987)
indicates that the galaxy must be reasonably compact, because there is no
reddening of component B relative to C.
Therefore, this possibility is unlikely to be true.
Another possibility is that the discrepancy may be due to a time delay
among the image components.
However, this may not also remedy the elliptical lens model, since the
predicted time delay (see tables 3 and 4) is too short compared with the
observed time variation of amplification (Vanderriest 1986).
The elliptical lens model may also raise other troubles.
It gives a rather
large amplification of undetected image D.
However, we can not rule out the possibility that component D
is so near to the lens that it suffers from absorption.

However, there is too much evidence to reject the elliptical
lens model.
Thus, as far as the lensing object is assumed to be isolated,
we are attracted to adopt a model that is qualitatively different from
the elliptical lens model.
A multipole-expansion model is the typical one,
which is obtained by expanding the Fermat potential in powers
of $1/|\mbox{\boldmath$x$}|$ in equation (\ref{Fermat}).
It should be remarked that the multipole expansion
model gives a better fit to the observations than does the elliptical model.
Here we quote the main result by the multipole expansion model
applied to PG 1115+080 (Kakigi et al. 1995; Fukuyama et al. 1997)

Here, the aspect of the lensing object effectively appears as the
multipole components as a whole.
We have thus also calculated the multipole components of the elliptical
lens model for comparison.
Here,
\begin{eqnarray}
m &\equiv& \frac{1}{\pi} \int d^2x' \kappa(\mbox{\boldmath$x'$})  \label{mm} \\
q_{ij} &\equiv& \frac{1}{\pi} \int d^2x' \kappa(\mbox{\boldmath$x'$}) (x'_i x'_j-
\frac{\vert \mbox{\boldmath$x'$} \vert^2}{2} \delta_{ij}) \qquad (i,j =1,2).
\label{qq}
\end{eqnarray}
In the elliptical lens model, if an image position lies
on the ellipse $b=b_0$ in equation (\ref{b2}),
it follows from equation (\ref{Newton}) that only the mass distribution
within $b=b_0$ can contribute to the deflection angle of this image
position.
Therefore, the multipole moments of the elliptical lens model
are estimated by integrating only the $b < b_0$ region in
equations (\ref{mm}) and (\ref{qq}).
Thus, each image has its own values of multipole moments.
The multipole-expansion model gives a monopole of $3.5 \times 10^{-11}$.
This value is dimensionless and $1 \times 10^{-10}$
corresponds to $4.4 \times 10^{11} h^{-1} \MO$.
Our elliptical lens models give a monopole of $4.4 - 4.7 \times 10^{-11}$
with respect to the outer-most image C and of $2.1\times 10^{-11}$
with respect to the inner-most image B, irrespective of the parameters.
These values are very near to those of the multipole-expansion model.
That is, the mass within the images is the same in the various models.
As for the quadrupole moment,
that of the elliptical lens model becomes large
as $e \sin{\gamma}$ increases, but there is no difference
for the cases $k=2$ and $k=3$.
Their values are about $1 \times 10^{-23}$, and
ten-times smaller than that of the multipole-expansion model.

The multipole-expansion model also asserts that the lensing
object has a significant dipole moment:
\begin{equation}
d_i \equiv \frac{1}{\pi} \int d^2x' \kappa(\mbox{\boldmath$x'$}) x'_i.
\label{dd}
\end{equation}
An elliptical lens does not have a dipole moment unless
the position of the center of mass is shifted from the origin
of coordinates.
The axis of the dipole moment obtained in the multipole-expansion
model (table 6) is nearly parallel to the declination.
In order to incorporate the dipole moment in the elliptical lens
model, we shifted the origin of coordinates (center of mass)
relative to the image positions
along the declination and tested
whether the data fitting is improved or not.
The result was negative.

The discussion so far developed is not advantageous to the single elliptical
lens model.
A decisive answer, however, requires more precise observations.
Recently, the precise spectrum of PG 1115+080 was obtained by
Michalitsianos et al. (1996).
They revealed that this QSO has a broad absorption line (BAL) of O
{\scriptsize VI}.
This means that this QSO has a large-scale outflow gas intrinsic to
this QSO.
These internal structures of the source must be reflected to that of the image
components, and increase the amount of information to be adjusted.
Here, as a preliminary step to this new stage, we have listed the result
of deformation assuming that the source is spherical.
Detailed formulation concerning the shape of the image was developed in
Kakigi et al. (1995).

Lastly, we comment on recent studies.
A closely connected paper by Keeton and Kochanek (1997) on
PG 1115+080 appeared in the Preprint Server
three months after our original paper.
Firstly, by allowing the parameter freedom to move the
lens position from the center of the observed position,
they practiced parameter fitting by isolating models.
As a result, the $\chi^2$ of the image positions and the amplifications
can not be small and the assumed lens
position goes beyond the error bar.
Their core radius is smaller than ours, since they moved
the lens position, but it was still larger than that of an
ordinary galaxy.
They then applied models corresponding to our
$k=2, 3, 4$ and de-Vaucouleurs model, all adopting the
perturbation by a group of galaxies $14^{\prime \prime}.5$
apart from the lensing galaxy,
and succeeded in obtaining $\chi^2/{\rm DOF} <1$
for all models.

Using the same weight as Keeton and Kochanek, a multipole-expansion
model gives
$\chi^2/{\rm DOF} \simeq 1.5$ (Kakigi 1997, private communication).
This value is smaller than those of Schechter et al. (1997),
whose models take the effect of nearby galaxies into
consideration by shear.
Therefore, the fact that an isolated transparent lens model
gives no good fitting does not necessarily lead us to
conclude the presence of a perturbation due to nearby galaxies.

Thus, $\chi^2/{\rm DOF}$ of the image positions and the
amplifications can be equal or less than unit in
several lens models.
For determining models,
a $\chi^2$ fitting is not sufficient.
Both quantitative and qualitative arguments from
various viewpoints complement this deficit.
For that purpose, analytical surveys of lensing models
like those developed in section two seem
to become more important.

This paper is summarized as follows.
We applied the elliptical lens model with softened power-law behaviors
of $k=2$ and $k=3$ to the multiple QSO PG 1115+080.
Using the obtained fitting parameters,
we calculated other observables.
Firstly, the elliptical lens models do not well produce
the image positions.
Secondly, the image amplifications do not well coincide
with the observed values.
Finally, we obtained a relatively large core radius.
For the above reasons,
we are forced to conclude that this model with $k=2$ and $k=3$
can not be consistent with the observations.
As an alternative model, if we accept a non-transparent model,
the lens may have an asymmetric
mass distribution with a non zero dipole moment, as was suggested by the
multipole-expansion model.
It is very natural to consider that this dipole moment reflects the presence
of the nearby group of galaxies (Young et al. 1981).
However, the direction of the dipole moment deviates from that of the
nearby group by $100^{\circ}$.
We have no clear explanation for this fact.

In any case, we have not derived any definite picture of the lensing
galaxy due to gravitational lensing.
We are enthusiastically waiting for more precise data, such as
spectroscopy profiles of the individual image components
(Michalitsianos, Oliversen 1995), their time variations and the image substructures
etc.
\vspace{1pc}\par

We are grateful to a referee who
made very constructive and
useful comments to us.
\vfill\eject


\section*{References}
\small

\re
	Angonin-Willaime M.-C., Hammer F., Rigaut F.\
	1993, in Proc. 31st Li\`ege Int. Astrophys. Colloq.,
	Gravitational Lenses in the Universe, ed J.Surdej,
	D.Fraipont-Caro, E.Gosset, S.Refsdal, M.Remy
	(Institut d'Astrophysique, Li\`ege) p85

\re
	Bahcall N.A., Lubin L.M.\ 1994,
	ApJ\  426, 513

\re
	Bar-Kana R.\ 1997,
	ApJ\  489, 21

\re
	Bertin G., Pignatelli E., Saglia R.P. \ 1993,
	A\&A\ 271, 381
\re
	Binney J., Tremaine S.\ 1987, Galactic Dynamics
	(Princeton University Press, Princeton) ch1

\re
	Bourassa R.R., Kantowski R., Norton T.D.\ 1973,
	ApJ\ 185, 747

\re
	Bourassa R.R., Kantowski R.\ 1975,
	ApJ\ 195, 13

\re
	Christian C.A., Crabtree D., Waddell P.\ 1987,
	ApJ\ 312, 45 

\re
	Fukuyama T., Kakigi Y., Okamura T.\ 1997,
	Int. J. Mod. Phys. D\ 6, 425

\re
	Grogin N.A., Narayan R.\ 1996,
	ApJ\ 464, 92

\re
	Kakigi Y., Okamura T., Fukuyama T.\ 1995,
	Int. J. Mod. Phys. D\ 4, 685 

\re
	Kassiola A., Kovner I.\ 1993,
	ApJ\ 417, 450

\re
	Keeton C.R., Kochanek C.S.\ 1997,
	ApJ\ 487, 42

\re
	Kent S.M., Falco E.E. \ 1988,
	AJ\ 96, 1570

\re
	Kormann R., Schneider P., Bartelmann M.\ 1994,
	A\&A\ 284, 285

\re
	Kristian J., Groth E.J., Shaya E.J., Schneider D.P., Holtzman J.A.,
	Baum W.A., Campbell B., Code A. et al.\ 1993,
	AJ\ 106, 1330

\re
	Lauer T.R.\ 1985,
	ApJ\ 292, 104

\re
	Lauer T.R., Faber S.M., Holtzman J.A., Baum W.A., Currie D.G.,
	Ewald S.P., Groth E.J., Hester J.J. et al.\ 1991,
	ApJ\ 369, L41

\re
	Michalitsianos A.G., Oliversen R.J. 1995,
	ApJ\ 439, 599

\re
	Michalitsianos A.G., Oliversen R.J., Nichols J.\ 1996,
	ApJ\ 461, 593

\re
	Miralda-Escud\'e J.\ 1993,
	ApJ\ 403, 497

\re
	Narasimha D., Subramanian K., Chitre S.M.\ 1982, 
	MNRAS\ 200, 941 

\re
	Schechter P.L.\
	1996, in Proc. Astrophysical Applications of Gravitational Lensing,
	ed C.S.Kochanek, J.N.Hewitt (Kluwer, Dordrecht) p263

\re
	Schechter P.L., Bailyn C.D., Barr R., Barvainis R.,
	Becker C.M., Bernstein G.M., Blakeslee J.P., Bus S.J. et al.\
	1997, ApJ\ 475, L85

\re
	Schramm T.\ 1990,
	A\&A\ 231, 19

\re
	Vanderriest C., Wl\'eric G.,Leli\`evre G., Schneider J.,
	Sol H.,Horville D., Renard L.,Servan B.\ 1986,
	A\&A\ 159, L5

\re
	Yoshida H., Omote M.\ 1988,
	Prog. Theor. Phys.\ 79, 1095 

\re
	Young P., Deverill R.S., Gunn J.E., Westphal J.A., Kristian J.\
	1981, ApJ\ 244, 723

\vfill\eject

\begin{center}
{\bf Figure Captions}
\end{center}

\begin{description}
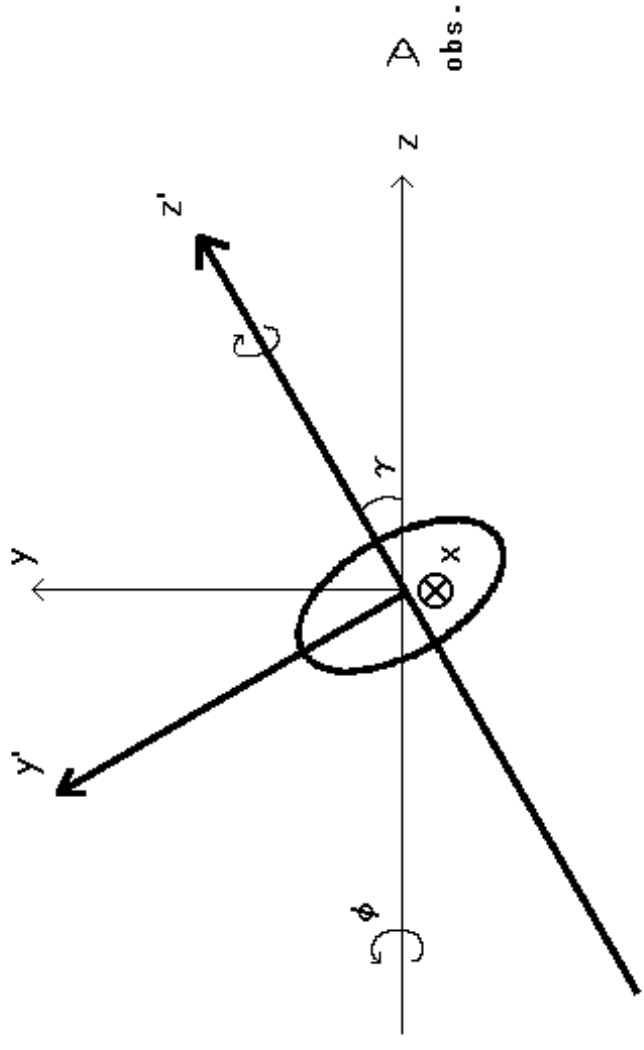
\item[Fig. 1.] Two Euler angles $\gamma$ and
$\phi$ in a spheroid.
$\phi$ is the rotation angle between the $x$-axis and the major axis
of the lens in the lens plane.

\item[Fig. 2.] Schematic diagram of the gravitational-lens geometry.
The solid line shows the real light path.

\item[Fig. 3.] Observed image positions (circles) and their numerical
calculations (squares).
The solid lines are the caustics and dashed lines
are the critical lines.
Here, the adopted parameters are those of $k$=3, $n$=10 and
$e \sin{\gamma}=0.5$ in table 4.

\end{description}

\newpage

\begin{table*}[h]
\begin{center}
Table~1. \hspace{4pt}Parameters of lensing object.
\end{center}
\vspace{6pt}
\begin{tabular*}{\textwidth}{@{\hspace{\tabcolsep}
\extracolsep{\fill}}p{6pc}ll}
\hline\hline\\[-6pt]
Parameters & Symbol \\[4pt]\hline\\[-6pt]
Eccentricity & $e$ \\
Inclination & $\sin{\gamma}$ \\
Core radius & $r_{\rm c}$ \\
Cut-off radius & $n\cdot r_{\rm c}$ \\
Source position & ($x_{\rm s},y_{\rm s}$) \\
Normalized central density & $\kappa_0$ \\
Rotation angle in lens plane & $\phi$ \\ \hline
\end{tabular*}
\vspace{6pt}\par\noindent
$*$ $e$ and $\sin{\gamma}$
appear only in the combined form of $e \sin{\gamma}$.
\end{table*}

\begin{table*}[h]
\begin{center}
Table~2. \hspace{4pt}Observed image positions and their amplifications
relative to C component in PG 1115+080.
\end{center}
\vspace{6pt}
\begin{tabular*}{\textwidth}{@{\hspace{\tabcolsep}
\extracolsep{\fill}}p{6pc}lccc}
\hline\hline\\[-6pt]
Image & $x_i$ & $y_i$ & $\mu/\mu_{\rm C}$ \\ \hline 
${\rm A}_1$ & $-0^{\prime \prime}.94$ & $-0^{\prime \prime}.73$ & 3.22 \\
\hline
${\rm A}_2$ & $-1^{\prime \prime}.11$ & $-0^{\prime \prime}.27$ & 2.49 \\
\hline
B & $0^{\prime \prime}.72$ & $-0^{\prime \prime}.60$ & 0.64 \\ \hline
C & $0^{\prime \prime}.33$ & $1^{\prime \prime}.35$ & 1.00 \\ \hline
\end{tabular*}
\end{table*}

\newpage

\begin{center}
Table~3. \hspace{4pt}Numerical calculations of the image
positions for the k=2 case.
\vspace{6pt}
\renewcommand{\arraystretch}{0.5}
\begin{tabular}{lccc}
\hline
$n$ &\multicolumn{3}{c}{5} \\ \hline
$e \sin{\gamma}$ & 0.3 & 0.4 &
0.5 \\ \hline
$r_{\rm c}$ ($h^{-1}$kpc)&$ 1^{\prime \prime}.86$ (5.16) & 
$1^{\prime \prime}.31$ (3.64) &
$0^{\prime \prime}.97$ (2.69) \\ \hline
$\kappa_0 \sqrt{1-e^2}$ ($h^2$ g ${\rm cm^{-3}}$) &
0.77 ($2.17\times10^{-23}$) &
0.82 ($3.28\times10^{-23}$) &
0.88 ($4.76\times10^{-23}$) \\ \hline
Total mass & $1.99\times10^{12}\MO h^{-1}$ &
$1.06\times10^{12}\MO h^{-1}$ &
$6.19\times10^{11}\MO h^{-1}$ \\ \hline
Source position & $(0^{\prime \prime}.014,-0^{\prime \prime}.020)$ &
$(0^{\prime \prime}.025,-0^{\prime \prime}.036)$ &
$(0^{\prime \prime}.039,-0^{\prime \prime}.055)$ \\ \hline
\multicolumn{4}{c}{\bf Image positions (Time delay lagging behind
C in day $h^{-1}$)} \\ \hline
${\rm A}_1$(fitting) & $(0^{\prime \prime}.52,1^{\prime \prime}.08) (1.5)$ &
$(0^{\prime \prime}.54,1^{\prime \prime}.06) (2.7)$ &
$(0^{\prime \prime}.54,1^{\prime \prime}.05) (4.1)$ \\
${\rm A}_1$(observed) & $(0^{\prime \prime}.59,1^{\prime \prime}.03)$ &
$(0^{\prime \prime}.58,1^{\prime \prime}.04)$ &
$(0^{\prime \prime}.57,1^{\prime \prime}.04)$ \\ \hline
${\rm A}_2$(fitting) & $(0^{\prime \prime}.94,0^{\prime \prime}.64) (1.5)$ &
$(0^{\prime \prime}.96,0^{\prime \prime}.62) (2.8)$ &
$(0^{\prime \prime}.99,0^{\prime \prime}.58) (4.2)$ \\
${\rm A}_2$(observed) & $(0^{\prime \prime}.92,0^{\prime \prime}.67)$ &
$(0^{\prime \prime}.92,0^{\prime \prime}.68)$ &
$(0^{\prime \prime}.91,0^{\prime \prime}.69)$ \\ \hline
B(fitting) & $(-0^{\prime \prime}.88,0^{\prime \prime}.32) (2.2)$ &
$(-0^{\prime \prime}.88,0^{\prime \prime}.31) (4.3)$ &
$(-0^{\prime \prime}.89,0^{\prime \prime}.29) (6.5)$ \\
B(observed) & $(-0^{\prime \prime}.89,0^{\prime \prime}.28)$ &
$(-0^{\prime \prime}.90,0^{\prime \prime}.27)$ &
$(-0^{\prime \prime}.90,0^{\prime \prime}.26)$ \\ \hline
C(fitting) & $(0^{\prime \prime}.23,-1^{\prime \prime}.37)$ &
$(0^{\prime \prime}.26,-1^{\prime \prime}.36)$ &
$(0^{\prime \prime}.28,-1^{\prime \prime}.36)$ \\
C(observed) & $(0^{\prime \prime}.21,-1^{\prime \prime}.37)$ &
$(0^{\prime \prime}.22,-1^{\prime \prime}.37)$ &
$(0^{\prime \prime}.24,-1^{\prime \prime}.37)$ \\ \hline
D(fitting) & $(-0^{\prime \prime}.20,0^{\prime \prime}.27)$ &
$(-0^{\prime \prime}.20,0^{\prime \prime}.09)$ &
$(-0^{\prime \prime}.18,0^{\prime \prime}.08)$ \\ \hline
\end{tabular}
\\
\begin{tabular}{lccc}
\hline 
$n$ &\multicolumn{3}{c}{10} \\ \hline
$e \sin{\gamma}$ & 0.3 & 0.4 &
0.5 \\ \hline
$r_{\rm c}$ ($h^{-1}$kpc)&$ 1^{\prime \prime}.69$ (4.70) & 
$1^{\prime \prime}.23$ (3.41) &
$0^{\prime \prime}.88$ (2.44) \\ \hline
$\kappa_0 \sqrt{1-e^2}$ ($h^2$ g ${\rm cm^{-3}}$) &
0.73 ($2.26\times10^{-23}$) &
0.77 ($3.28\times10^{-23}$) &
0.84 ($5.00\times10^{-23}$) \\ \hline
Total mass & $3.67\times10^{12}\MO h^{-1}$ &
$2.05\times10^{12}\MO h^{-1}$ &
$1.15\times10^{12}\MO h^{-1}$ \\ \hline
Source position & $(0^{\prime \prime}.014,-0^{\prime \prime}.021)$ &
$(0{\prime \prime}.025^,-0^{\prime \prime}.035)$ &
$(0^{\prime \prime}.040,-0^{\prime \prime}.054)$ \\ \hline
\multicolumn{4}{c}{\bf Image positions (Time delay lagging behind
C in day $h^{-1}$)} \\ \hline
${\rm A}_1$(fitting) & $(0^{\prime \prime}.56,1^{\prime \prime}.04) (1.5)$ &
$(0^{\prime \prime}.52,1^{\prime \prime}.07) (2.7)$ &
$(0^{\prime \prime}.54,1^{\prime \prime}.05) (4.1)$ \\
${\rm A}_1$(observed) & $(0^{\prime \prime}.58,1^{\prime \prime}.04)$ &
$(0^{\prime \prime}.58,1^{\prime \prime}.04)$ &
$(0^{\prime \prime}.57,1^{\prime \prime}.05)$ \\ \hline
${\rm A}_2$(fitting) & $(0^{\prime \prime}.92,0^{\prime \prime}.67) (1.6)$ &
$(0^{\prime \prime}.92,0^{\prime \prime}.61) (2.8)$ &
$(0^{\prime \prime}.99,0^{\prime \prime}.57) (4.2)$ \\
${\rm A}_2$(observed) & $(0^{\prime \prime}.91,0^{\prime \prime}.69)$ &
$(0^{\prime \prime}.92,0^{\prime \prime}.68)$ &
$(0^{\prime \prime}.91,0^{\prime \prime}.70)$ \\ \hline
B(fitting) & $(-0^{\prime \prime}.90,0^{\prime \prime}.31) (2.4)$ &
$(-0^{\prime \prime}.89,0^{\prime \prime}.30) (4.3)$ &
$(-0^{\prime \prime}.89,0^{\prime \prime}.29) (6.6)$ \\
B(observed) & $(-0^{\prime \prime}.90,0^{\prime \prime}.27)$ &
$(-0^{\prime \prime}.90,0^{\prime \prime}.27)$ &
$(-0^{\prime \prime}.90,0^{\prime \prime}.26)$ \\ \hline
C(fitting) & $(0^{\prime \prime}.23,-1^{\prime \prime}.35)$ &
$(0^{\prime \prime}.26,-1^{\prime \prime}.36)$ &
$(0^{\prime \prime}.28,-1^{\prime \prime}.36)$ \\
C(observed) & $(0^{\prime \prime}.23,-1^{\prime \prime}.37)$ &
$(0^{\prime \prime}.23,-1^{\prime \prime}.37)$ &
$(0^{\prime \prime}.24,-1^{\prime \prime}.37)$ \\ \hline
D(fitting) & $(-0^{\prime \prime}.30,0^{\prime \prime}.36)$ &
$(-0^{\prime \prime}.24,0^{\prime \prime}.29)$ &
$(-0^{\prime \prime}.17,0^{\prime \prime}.22)$ \\ \hline
\end{tabular}\\
\vspace{6pt}
$*$
The upper part is the adopted values of the parameters,
and the lower part is the
corresponding theoretical values and time delay in units of days $h^{-1}$.
We have also listed the observed values for a comparison.
The parenthesized values in lines of $r_{\rm c}$ and $\kappa_0 \sqrt{1-e^2}$
are the physical core size in units of $h^{-1}$kpc and the central density,
$\rho_0 \sqrt{1-e^2}$ in units of $h^2$ g ${\rm cm^{-3}}$, respectively.
Here, we assume $\Omega_0=1$, $H_0=100h$ km ${\rm s^{-1}}$
Mpc, $z_{\rm l}=0.3$ and
$z_{\rm s}=1.72$.
\end{center}

\newpage
\begin{center}
Table~4. \hspace{4pt}The same calculations as in table 3 for the $k$=3 case.
\vspace{6pt}
\renewcommand{\arraystretch}{0.5}
\begin{tabular}{lccc}
\hline 
$n$ &\multicolumn{3}{c}{5} \\ \hline
$e \sin{\gamma}$ & 0.4 & 0.5 &
0.6 \\ \hline
$r_{\rm c}$ ($h^{-1}$kpc)&$ 1^{\prime \prime}.90$ (5.27) & 
$1^{\prime \prime}.48$ (4.11) &
$1^{\prime \prime}.19$ (3.30) \\ \hline
$\kappa_0 \sqrt{1-e^2}$ ($h^2$ g ${\rm cm^{-3}}$) &
1.13 ($3.12\times10^{-23}$) &
1.19 ($4.21\times10^{-23}$) &
1.28 ($5.64\times10^{-23}$) \\ \hline
Total mass & $1.12\times10^{12}\MO h^{-1}$ &
$7.17\times10^{11}\MO h^{-1}$ &
$4.99\times10^{11}\MO h^{-1}$ \\ \hline
Source position & $(0^{\prime \prime}.024,-0^{\prime \prime}.037)$ &
$(0^{\prime \prime}.038,-0^{\prime \prime}.058)$ &
$(0^{\prime \prime}.057,-0^{\prime \prime}.085)$ \\ \hline
\multicolumn{4}{c}{\bf Image positions (Time delay lagging behind
C in day $h^{-1}$)} \\ \hline
${\rm A}_1$(fitting) & $(0^{\prime \prime}.51,1^{\prime \prime}.08) (2.8)$ &
$(0^{\prime \prime}.53,1^{\prime \prime}.06) (4.3)$ &
$(0^{\prime \prime}.53,1^{\prime \prime}.05) (6.3)$ \\
${\rm A}_1$(observed) & $(0^{\prime \prime}.59,1^{\prime \prime}.03)$ &
$(0^{\prime \prime}.58,1^{\prime \prime}.04)$ &
$(0^{\prime \prime}.57,1^{\prime \prime}.04)$ \\ \hline
${\rm A}_2$(fitting) & $(0^{\prime \prime}.95,0^{\prime \prime}.63) (2.9)$ &
$(0^{\prime \prime}.97,0^{\prime \prime}.61) (4.4)$ &
$(1^{\prime \prime}.00,0^{\prime \prime}.57) (6.4)$ \\
${\rm A}_2$(observed) & $(0^{\prime \prime}.92,0^{\prime \prime}.67)$ &
$(0^{\prime \prime}.92,0^{\prime \prime}.68)$ &
$(0^{\prime \prime}.91,0^{\prime \prime}.69)$ \\ \hline
B(fitting) & $(-0^{\prime \prime}.88,0^{\prime \prime}.31) (4.2)$ &
$(-0^{\prime \prime}.89,0^{\prime \prime}.30) (6.6)$ &
$(-0^{\prime \prime}.89,0^{\prime \prime}.28) (9.7)$ \\
B(observed) & $(-0^{\prime \prime}.89,0^{\prime \prime}.28)$ &
$(-0^{\prime \prime}.90,0^{\prime \prime}.27)$ &
$(-0^{\prime \prime}.90,0^{\prime \prime}.26)$ \\ \hline
C(fitting) & $(0^{\prime \prime}.23,-1^{\prime \prime}.37)$ &
$(0^{\prime \prime}.24,-1^{\prime \prime}.37)$ &
$(0^{\prime \prime}.26,-1^{\prime \prime}.37)$ \\
C(observed) & $(0^{\prime \prime}.21,-1^{\prime \prime}.37)$ &
$(0^{\prime \prime}.22,-1^{\prime \prime}.37)$ &
$(0^{\prime \prime}.24,-1^{\prime \prime}.37)$ \\ \hline
D(fitting) & $(-0^{\prime \prime}.17,0^{\prime \prime}.15)$ &
$(-0^{\prime \prime}.16,0^{\prime \prime}.13)$ &
$(-0^{\prime \prime}.15,0^{\prime \prime}.12)$ \\ \hline
\end{tabular}
\\
\begin{tabular}{lccc}
\hline 
$n$ &\multicolumn{3}{c}{10} \\ \hline
$e \sin{\gamma}$ & 0.4 & 0.5 &
0.6 \\ \hline
$r_{\rm c}$ ($h^{-1}$kpc)&$ 1^{\prime \prime}.88$ (5.22) & 
$1^{\prime \prime}.46$ (4.05) &
$1^{\prime \prime}.16$ (3.22) \\ \hline
$\kappa_0 \sqrt{1-e^2}$ ($h^2$ g ${\rm cm^{-3}}$) &
1.11 ($3.10\times10^{-23}$) &
1.18 ($4.24\times10^{-23}$) &
1.27 ($5.74\times10^{-23}$) \\ \hline
Total mass & $1.62\times10^{12}\MO h^{-1}$ &
$1.04\times10^{12}\MO h^{-1}$ &
$7.07\times10^{11}\MO h^{-1}$ \\ \hline
Source position & $(0^{\prime \prime}.024,-0^{\prime \prime}.037)$ &
$(0^{\prime \prime}.039,-0^{\prime \prime}.058)$ &
$(0^{\prime \prime}.057,-0^{\prime \prime}.084)$ \\ \hline
\multicolumn{4}{c}{\bf Image positions (Time delay lagging behind
C in day $h^{-1}$)} \\ \hline
${\rm A}_1$(fitting) & $(0^{\prime \prime}.51,1^{\prime \prime}.08) (2.7)$ &
$(0^{\prime \prime}.53,1^{\prime \prime}.06) (4.3)$ &
$(0^{\prime \prime}.53,1^{\prime \prime}.05) (6.2)$ \\
${\rm A}_1$(observed) & $(0^{\prime \prime}.59,1^{\prime \prime}.03)$ &
$(0^{\prime \prime}.58,1^{\prime \prime}.04)$ &
$(0^{\prime \prime}.57,1^{\prime \prime}.05)$ \\ \hline
${\rm A}_2$(fitting) & $(0^{\prime \prime}.96,0^{\prime \prime}.62) (2.8)$ &
$(0^{\prime \prime}.97,0^{\prime \prime}.61) (4.4)$ &
$(1^{\prime \prime}.00,0^{\prime \prime}.57) (6.4)$ \\
${\rm A}_2$(observed) & $(0^{\prime \prime}.93,0^{\prime \prime}.67)$ &
$(0^{\prime \prime}.92,0^{\prime \prime}.68)$ &
$(0^{\prime \prime}.91,0^{\prime \prime}.69)$ \\ \hline
B(fitting) & $(-0^{\prime \prime}.89,0^{\prime \prime}.31) (4.1)$ &
$(-0^{\prime \prime}.89,0^{\prime \prime}.30) (6.7)$ &
$(-0^{\prime \prime}.89,0^{\prime \prime}.28) (9.7)$ \\
B(observed) & $(-0^{\prime \prime}.90,0^{\prime \prime}.28)$ &
$(-0^{\prime \prime}.90,0^{\prime \prime}.27)$ &
$(-0^{\prime \prime}.90,0^{\prime \prime}.26)$ \\ \hline
C(fitting) & $(0^{\prime \prime}.24,-1^{\prime \prime}.37)$ &
$(0^{\prime \prime}.25,-1^{\prime \prime}.37)$ &
$(0^{\prime \prime}.26,-1^{\prime \prime}.37)$ \\
C(observed) & $(0^{\prime \prime}.21,-1^{\prime \prime}.37)$ &
$(0^{\prime \prime}.22,-1^{\prime \prime}.37)$ &
$(0^{\prime \prime}.24,-1^{\prime \prime}.37)$ \\ \hline
D(fitting) & $(-0^{\prime \prime}.17,0^{\prime \prime}.15)$ &
$(-0^{\prime \prime}.16,0^{\prime \prime}.13)$ &
$(-0^{\prime \prime}.15,0^{\prime \prime}.12)$ \\ \hline
\end{tabular}
\end{center}

\begin{center}
Table~5. \hspace{4pt}Numerical calculations of image amplifications.
\vspace{6pt}
\renewcommand{\arraystretch}{0.5}
\begin{tabular}{lcccccc|cccccc}
\hline
$k$ & \multicolumn{6}{c|}{2} & \multicolumn{6}{c}{3} \\ \hline 
$e \sin{\gamma}$ & \multicolumn{2}{c}{0.3} & \multicolumn{2}{c}{0.4} &
\multicolumn{2}{c|}{0.5} & \multicolumn{2}{c}{0.4} &
\multicolumn{2}{c}{0.5} & \multicolumn{2}{c}{0.6} \\ \hline
$n$ & 5 & 10 & 5 & 10 & 5 & 10 & 5 & 10 & 5 & 10 & 5 & 10 \\ \hline
${\rm A}_1/$C & 4.45 & 6.15 & 3.92 & 3.65 & 3.03 & 2.91 & 4.17 &
4.24 & 4.00 & 4.02 & 3.49 & 3.45 \\ \hline
${\rm A}_2/$C & 4.45 & 4.20 & 3.81 & 3.61 & 2.91 & 2.74 & 4.38 & 4.20 &
3.96 & 3.89 & 3.29 & 3.28 \\ \hline
B/C & 2.31 & 1.95 & 1.91 & 1.82 & 1.54 & 1.44 & 2.22 & 2.16 &
1.90 & 1.86 & 1.57 & 1.53  \\ \hline
D/C & 1.90 & 2.67 & 1.08 & 1.67 & 0.72 & 0.85 & 1.37 & 1.34 &
1.03 & 1.00 & 0.73 & 0.70  \\ \hline
\end{tabular}
\end{center}

\begin{center}
Table~6. \hspace{4pt}Numerical calculations by the multipole-expansion
model. 
\vspace{6pt}
\renewcommand{\arraystretch}{0.5}
\begin{tabular}{lccc}
\hline
m & \multicolumn{3}{c}{$3.50 \times 10^{-11}$} \\ \hline
$|d|$ & \multicolumn{3}{c}{$3.77 \times 10^{-17} (7.80 \times 10^{-19}, 3.77 \times 10^{-17})$} \\
\hline
$|q|$ & \multicolumn{3}{c}{$1.07 \times 10^{-22} (-6.93 \times 10^{-23}, 8.10 \times 10^{-23})$} \\
\hline
   &  {\bf Image positions} & {\bf Amplification relative to C}
& {\bf Time delay lagging behind C in day $h^{-1}$} \\ \hline
${\rm A}_1$ & ($-0.9383^{\prime \prime}, -0.7317^{\prime \prime}$) &
3.113 & 9.757 \\ \hline
${\rm A}_2$ & ($-1.110^{\prime \prime}, -0.2748^{\prime \prime}$) &
2.892 & 11.448 \\ \hline
B & ($0.7205^{\prime \prime}, -0.6004^{\prime \prime}$) &
0.7467 & 19.407 \\ \hline
C & ($0.3294^{\prime \prime}, 1.350^{\prime \prime}$) &
1.000 & 0.000 \\ \hline
D & ($0.5915^{\prime \prime}, -0.4450^{\prime \prime}$) &
0.1780 & 19.225 \\ \hline
E & ($-0.4494^{\prime \prime}, 0.08641^{\prime \prime}$) &
0.005871 & 3.076 \\ \hline
\end{tabular}
\end{center}
$*$ Multipole components are defined by equations (\ref{mm})-(\ref{dd}).
The bracket in lines of $|d|$ and $|q|$ means $(d_1, d_2)$ and $(q_{11},
q_{12})$ components, respectively.
(From Fukuyama et al. 1997)

\begin{table*}[h]
\begin{center}
Table~7. \hspace{4pt}Calculated image eccentricity and angle between
the major axis of the image and the $x$-axis for the $k$=2 case.
\end{center}
\vspace{6pt}
\begin{tabular*}{\textwidth}{@{\hspace{\tabcolsep}
\extracolsep{\fill}}p{6pc}lcccccc}
\hline\hline\\[-6pt]
$e \sin{\gamma}$ & \multicolumn{2}{c}{0.3} & \multicolumn{2}{c}{0.4} &
\multicolumn{2}{c}{0.5}  \\ \hline
$n$ & 5 & 10 & 5 & 10 & 5 & 10 \\ \hline
${\rm A}_1$ Eccentricity(Axis ratio)& 0.9958(0.092)&0.9981(0.062) &
0.9949(0.101) &0.9944(0.106) &0.9921(0.125) &0.9917(0.129) \\ 
Angle &$-30^{\circ}.2$  &$-32^{\circ}.7$ &$-30^{\circ}.7$  &$-29^{\circ}.6$&
$-30^{\circ}.2$ &$-30^{\circ}.2$ \\ \hline
${\rm A}_2$ Eccentricity(Axis ratio)&0.9944(0.106) &0.9952(0.098) &0.9931(0.117) &
0.9925(0.122) &0.9890(0.148) &0.9885(0.151)  \\ 
Angle &$-60^{\circ}.6$  &$-58^{\circ}.5$ &$-61^{\circ}.0$  &$-61^{\circ}.4$ &
$-62^{\circ}.3$ &$-62^{\circ}.7$ \\ \hline
B Eccentricity(Axis ratio)&0.8378(0.546) &0.8662(0.500) &0.8193(0.573) &
0.8389(0.544) &0.7995(0.601) &0.8175(0.576)  \\ 
Angle &$74^{\circ}.5$  &$74^{\circ}.8$ &$74^{\circ}.5$  &$75^{\circ}.1$ &
$75^{\circ}.3$ &$75^{\circ}.2$ \\ \hline
C Eccentricity(Axis ratio)&0.9756(0.220) &0.9787(0.205) &0.9735(0.229) &
0.9726(0.233) &0.9707(0.240) &0.9693(0.246)  \\ 
Angle &$10^{\circ}.9$  &$11^{\circ}.0$ &$12^{\circ}.2$  &$12^{\circ}.3$ &
$13^{\circ}.1$ &$13^{\circ}.2$ \\ \hline
\end{tabular*}
\end{table*}

\begin{table*}[h]
\begin{center}
Table~8. \hspace{4pt}The same calculations as in table 7 for the $k$=3 case.
\end{center}
\vspace{6pt}
\begin{tabular*}{\textwidth}{@{\hspace{\tabcolsep}
\extracolsep{\fill}}p{6pc}lcccccc}
\hline\hline\\[-6pt]
$e \sin{\gamma}$ & \multicolumn{2}{c}{0.4} &\multicolumn{2}{c}{0.5} &
 \multicolumn{2}{c}{0.6} \\ \hline
$n$ & 5 & 10 & 5 & 10 & 5 & 10 \\ \hline
${\rm A}_1$ Eccentricity(Axis ratio)& 0.9952(0.098) &0.9952(0.098) &0.9948(0.102) &
0.9948(0.102) &0.9934(0.115) &0.9933(0.116)  \\ 
Angle &$-28^{\circ}.7$ &$-28^{\circ}.6$ &$-28^{\circ}.9$ &$-28^{\circ}.9$ &
$-27^{\circ}.6$ &$-27^{\circ}.6$ \\ \hline
${\rm A}_2$ Eccentricity(Axis ratio)&0.9937(0.112) &0.9933(0.115) &0.9925(0.122) &
0.9922(0.124) &0.9890(0.148) &0.9892(0.146)  \\ 
Angle &$-60^{\circ}.5$ &$-61^{\circ}.1$ &$-60^{\circ}.6$ &$-60^{\circ}.6$ &
$-61^{\circ}.7$ &$-61^{\circ}.7$ \\ \hline
B Eccentricity(Axis ratio)&0.8025(0.597) &0.8202(0.572) &0.7860(0.618) &
0.7854(0.619) &0.7129(0.701) &0.7209(0.693)  \\ 
Angle &$74^{\circ}.6$ &$74^{\circ}.7$ &$74^{\circ}.6$ &$74^{\circ}.5$ &
$74^{\circ}.8$ &$74^{\circ}.8$ \\ \hline
C Eccentricity(Axis ratio)&0.9735(0.228) &0.9739(0.227) &0.9704(0.241) &
0.9704(0.242) &0.9659(0.259) &0.9656(0.260)  \\ 
Angle &$10^{\circ}.4$ &$10^{\circ}.9$ &$10^{\circ}.4$ &$10^{\circ}.8$ &
$10^{\circ}.6$ &$10^{\circ}.6$ \\ \hline
\end{tabular*}
\end{table*}

\end{document}